\documentclass[aps,twocolumn,showpacs,prl,superscriptaddress,unsortedaddress]{revtex4}
\usepackage{amssymb}
\usepackage{epsf}
\usepackage{graphicx}

\newcommand{\etal}{{\it et al.}}

\begin{document}

\title{Unusual Fermi surface nesting in parent compounds of iron
arsenic high temperature superconductors revealed by Angle Resolved
Photoemission Spectroscopy}
\author{Takeshi Kondo}
\affiliation{Ames Laboratory and Department of Physics and Astronomy, Iowa State
University, Ames, Iowa 50011, USA}
\author{R. M. Fernandes}
\affiliation{Ames Laboratory and Department of Physics and Astronomy, Iowa State
University, Ames, Iowa 50011, USA}
\author{R.~Khasanov}
\affiliation{Laboratory for Muon Spin Spectroscopy, Paul Scherrer Institut, CH-5232
Villigen PSI, Switzerland}
\author{Chang Liu}
\affiliation{Ames Laboratory and Department of Physics and Astronomy, Iowa State
University, Ames, Iowa 50011, USA}
\author{A. D. Palczewski}
\affiliation{Ames Laboratory and Department of Physics and Astronomy, Iowa State
University, Ames, Iowa 50011, USA}
\author{Ni Ni}
\affiliation{Ames Laboratory and Department of Physics and Astronomy, Iowa State
University, Ames, Iowa 50011, USA}
\author{M.~Shi}
\affiliation{Swiss Light Source, Paul Scherrer Institut, CH-5232 Villigen PSI, Switzerland}
\author{A. Bostwick}
\affiliation{Advanced Light Source, Berkeley National Laboratory, Berkeley, California
94720, USA}
\author{E. Rotenberg}
\affiliation{Advanced Light Source, Berkeley National Laboratory, Berkeley, California
94720, USA}
\author{J. Schmalian}
\affiliation{Ames Laboratory and Department of Physics and Astronomy, Iowa State
University, Ames, Iowa 50011, USA}
\author{S. L. Bud'ko}
\affiliation{Ames Laboratory and Department of Physics and Astronomy, Iowa State
University, Ames, Iowa 50011, USA}
\author{P. C. Canfield}
\affiliation{Ames Laboratory and Department of Physics and Astronomy, Iowa State
University, Ames, Iowa 50011, USA}
\author{A. Kaminski}
\affiliation{Ames Laboratory and Department of Physics and Astronomy, Iowa State
University, Ames, Iowa 50011, USA}
\date{\today}

\begin{abstract}
We use angle resolved photoemission spectroscopy (ARPES) to study the band
structure of BaFe$_2$As$_2$ and CaFe$_2$As$_2$, two of the parent compounds of the
iron arsenic high temperature superconductors. We find clear evidence for
band back folding and hybridization demonstrating that conduction electrons
are strongly affected by the emergence of magnetic order. Our high quality data revealed that although the Fermi surface is strongly three-dimensional, it does indeed have long parallel segments along the $k_z$ direction that can lead to
the emergence of magnetic order. More interestingly, we find very unusual incommensurate
nesting of the Fermi surface in the $a-b$ plane that is present only at low temperatures. 
We speculate that this is a signature of a failed Charge Density Wave (CDW) state that was predicted by renormalization group studies.
\end{abstract}

\pacs{79.60.-i, 74.25.Jb, 74.70.-b}
\maketitle
Iron arsenic superconductors \cite{Original} display a
fascinating interplay between magnetism and superconductivity. The influence
of the magnetism on the electronic properties and the role it plays in high
temperature superconductivity of these materials is a subject of a lively
debate within the condensed matter physics community. Neutron scattering
experiments\cite{Neutron,Goldman_lattice} reported antiferromagnetic (AF) ordering and 3D character of magnetic interactions\cite{Rob1} in non-doped parent compounds of iron arsenides. A number of previous
studies attributed the magnetic ordering to itinerant Spin Density Wave
(SDW) \cite{Feng, Hasan, Rob2}, while others favored magnetic order due to
localized moments \cite{Zhou}. The AF-ordering 
 is reminiscent of that of the cuprate high
temperature superconductors, which occurs in
the un-doped materials \cite{CuprateNeutron}. 
The AF ordering introduces a new
zone boundary (antiferromagnetic zone boundary - AFZB) leading to band
back folding effects \cite{ElectronCuprate1,ElectronCuprate2}. The main
difference between the iron arsenides and cuprates is that the parent (undoped)
phase in the former is not insulating but metallic \cite{Rotter,NiNi1,NiNi2,NiNi3}. This opens
the possibility of investigating the interplay between long range magnetic
ordering and the conduction electrons. Understanding this interaction is
essential to elucidate the origin of the fascinating properties displayed by
the new, iron-arsenic based superconductors, and may also help
to better understand the relation between magnetic and electronic
excitations in the cuprates where we do not have access to a conducting,
magnetically ordered state of the parent, undoped materials. In this letter, we present data from angle resolved
photoemission spectroscopy (ARPES) on the electronic properties of BaFe$_2$As$_2$
and CaFe$_2$As$_2$, two of the parent compounds for the iron arsenic high
temperature superconductors. We find strong band back folding and
hybridization effects that lead to a reconstruction of the Fermi surface (FS)
below the magnetic ordering temperature ($T_N$), which gives rise to
a characteristic ``flower-like" FS. We compare our data to itinerant spin model
calculations and find them in reasonable agreement. 
These signatures of the
magnetic ordering in the electronic structure vary significantly with value
of the $k_z$ momentum with coupling between charge carriers
and magnetism being strongest whenever an efficient
low energy scattering between the Fermi surface sheets
around two sets of Fermi surfaces is kinematically possible.
This clearly demonstrates the role of interband magnetic scattering in the iron arsenides.
We were able to identify long parallel segments along the $k_z$ direction
spanning nearly half of the Brillouin zone that may play an essential role in
the emergence of magnetic order in these materials. More interestingly, we
discovered long parallel segments of the Fermi surface within the $a-b$ plane. 
This new nesting vector is significantly shorter 
than the $(\pi/a, \pi/a)$ vector related
to the magnetic order. We speculate that these nested parts of the Fermi
surface may be a precursor of a failed CDW order predicted by 
renormalization group studies \cite{DHLEE,DHLEE2}.

\begin{figure}[tbp]
\includegraphics[width=3.in]{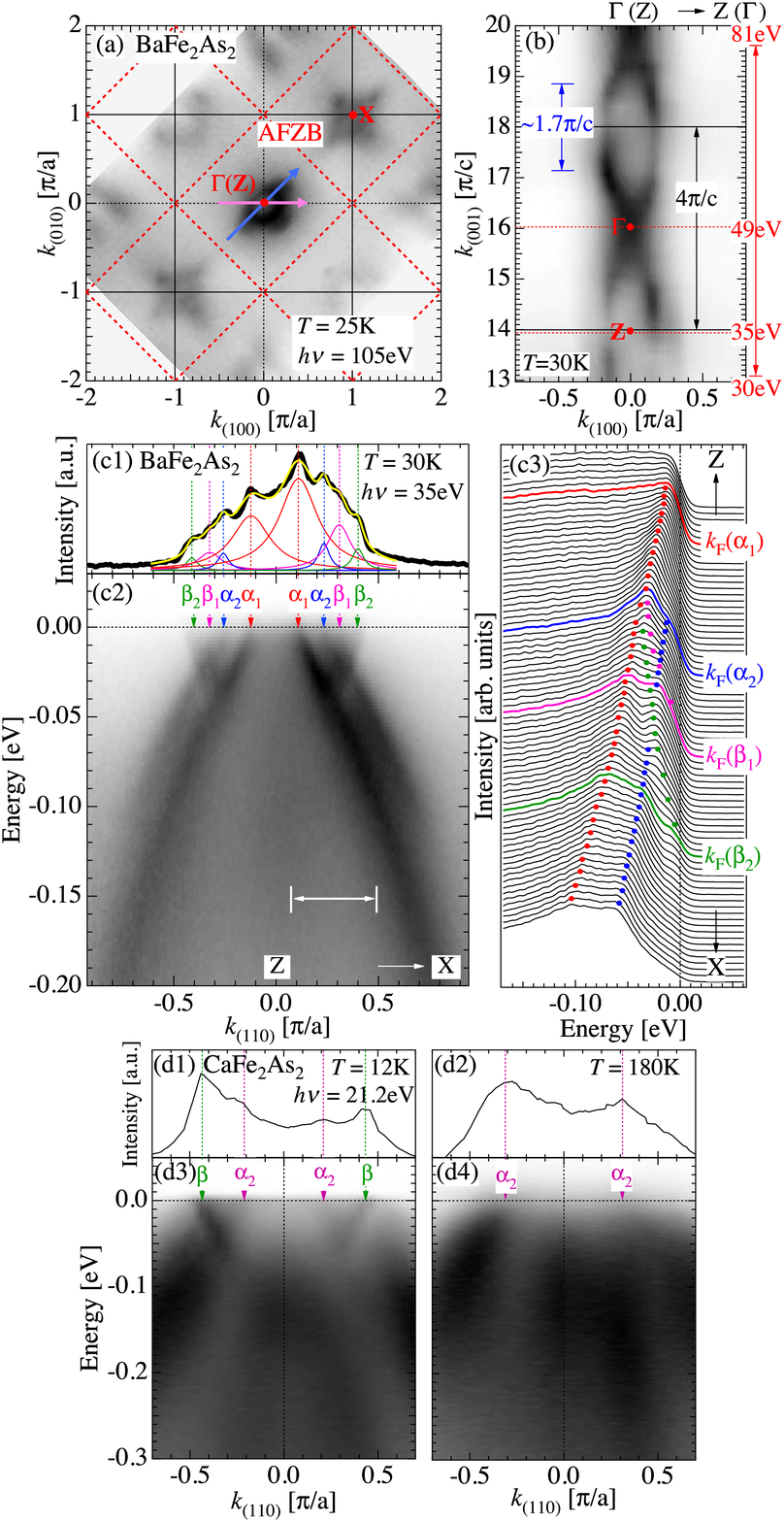}
\caption{(Color online) (a-c) BaFe$_2$As$_2$ and (d) CaFe$_2$As$_2$ data.
 FS map (a) along $a$-$b$ and (b) along $k_{
\mathrm{(001)}}$ and $k_{\mathrm{(100)}}$ (pink arrow in (a)). (Inner potential: $V_0  \equiv$ 13eV) Blue arrow indicates the parallel segment. (c1) MDC of (c2) at Fermi energy (black curve) and the fitting curve (yellow). (c2) Band dispersion map
along a diagonal cut (blue arrow in (a)). (c3) EDCs of (c2, a white arrow range).
Colored circles follow hole-bands ($\protect\alpha_1$ and
$\protect\alpha_2$) and electron-bands ($\protect\beta_1$ and $\protect\beta
_2$). (d1-d2) MDC at Fermi level of
(a3-b3). Band dispersion map along a diagonal cut (d3) below and (d4) above $T_N$. }
\label{fig1}
\end{figure}

\begin{figure}[tbp]
\includegraphics[width=2.8in]{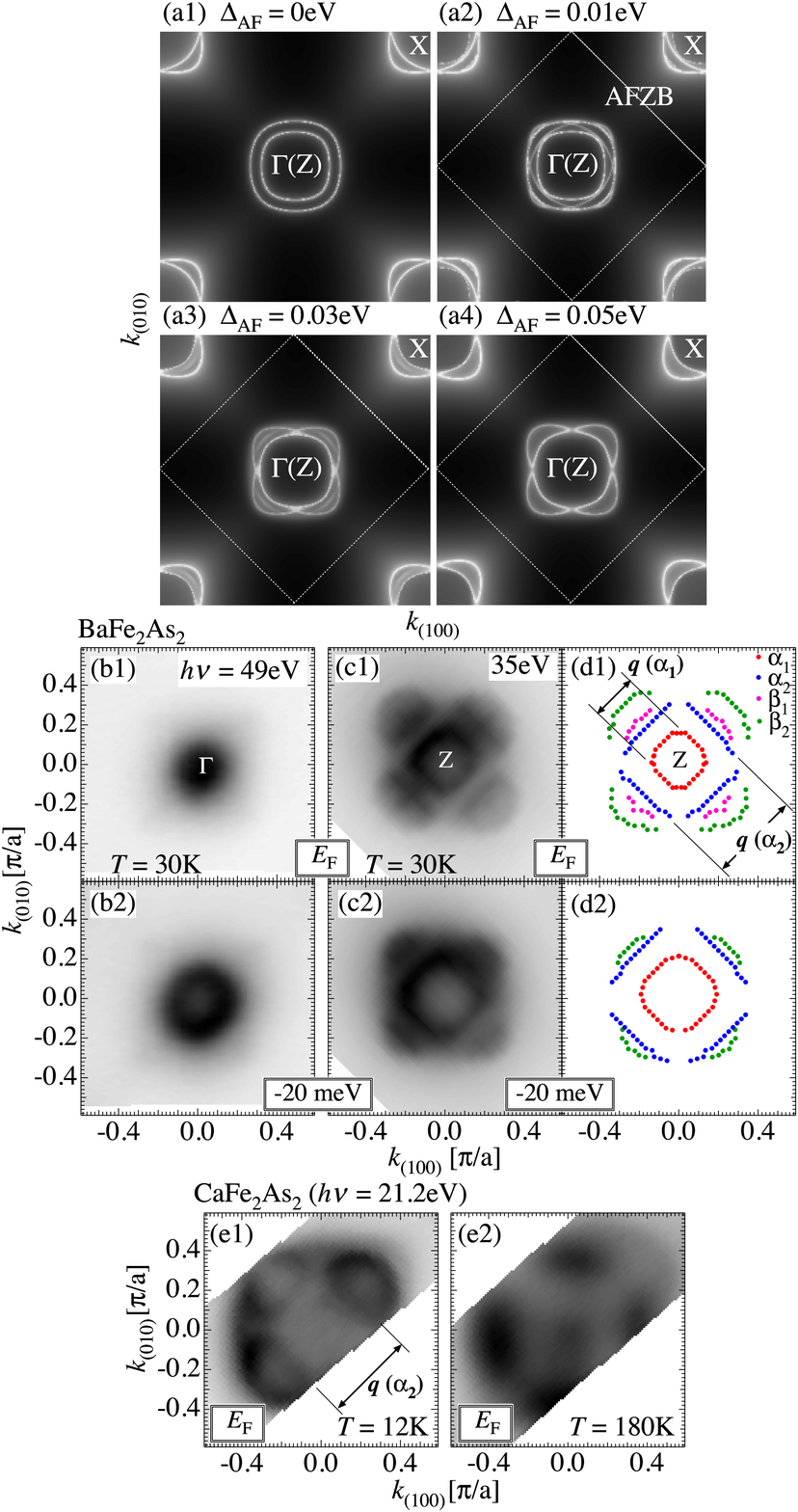}
\caption{(Color online) (a) Model calculation, (b-d) BaFe$_2$As$_2$, and (e) CaFe$_2$As$_2$ data. 
(a1) Model FSs in non-magnetic phase. Reconstructed FSs
 for $\Delta _{\mathrm{AF}}$ of (a2) 0.01 eV, (a3) 0.03 eV,
and (a4) 0.05 eV. ARPES intensity about
the Fermi energy  and $E=$-20meV 
obtained at (b1-b2) $h
\protect\nu=49$eV (near $\Gamma$) and (c1-c2) $h\protect\nu=35$eV (near Z).
(d1) FS and (d2) energy contour plots at $E=$-20meV for $h\protect\nu=49$eV data.
FS map close to the zone center along $a$-$b$ plane  (e1)
below and (e2) above $T_N$. Dimension arrows indicate nesting vector $q(\alpha_1)$=$0.08(\pi/a, \pi/a)$ and $q(\alpha_2)$=$0.32(\pi/a, \pi/a)$ for hole-band $\alpha_1$ and $\alpha_2$, respectively.
}
\label{fig2}
\end{figure}

Single crystals of BaFe$_{2}$As$_{2}$ and CaFe$_{2}$As$_{2}$ were grown out
of a FeAs flux as well as Sn flux using conventional high-temperature
solution growth techniques\cite{NiNi1,NiNi2,NiNi3}. BaFe$_{2}$As$_{2}$ and
CaFe$_{2}$As$_{2}$ undergo a tetragonal to orthorhombic structural
transition simultaneously with a paramagnetic to
antiferromagnetic transition below $T_{S}\simeq T_{N}\simeq $130K and 170K,
respectively\cite{Rotter, Goldman_lattice,NiNi2,NiNi3}. ARPES data were measured at
the SIS beamline of Swiss Light Source, Switzerland, the beamlines 7.0.1 of
the Advanced Light Source (ALS), USA, using a Scienta R4000 analyzer and
Ames Laboratory using a Scienta SES2002 with a Gammadata VUV5010 photon
source. Energy and angular resolutions were $10-30$meV and $\sim
$0.1$^{\circ }$, respectively. The high symmetry points X and Z
for both two phases are defined to be ($\pi/a$, $\pi/a$($b$), 0) and
(0, 0, $2\pi/c$), respectively, with $k_x$ ($k_{(100)}$) and $k_y$
($k_{(010)}$) axes along the Fe-As bonds.

Band calculations predict two types of nearly circular FSs in the
paramagnetic state: hole-like FSs centered at $\Gamma $(Z) and electron-like FSs
centered at X\cite{ChangPRL,BandCalculation}. Below the magnetic ordering
temperature one would expect back folding and hybridization of these bands
leading to a reconstruction of the FS\cite{BandCalculation}. Figure 1 (a)
shows a typical FS map from BaFe$_{2}$As$_{2}$ sample measured
with 105 eV photons deep in the magnetically ordered state. Dark areas indicate the locations of the
Fermi surfaces (FSs), which indeed display a more complicated structure 
than the one predicted in the paramagnetic state. 
To examine the character of the FSs, we show in Fig.1 (c2) an ARPES intensity  along a diagonal direction (blue arrow in Fig.1 (a)) measured with 35eV photons. 
In addition to hole-like bands that are
characteristic of the paramagnetic state, we observe electron-like bands that
are back folded from X-points about the antiferromagnetic zone boundary (AFZB, marked as red dashed lines in Fig.1(a)). 
The presence of multiple Fermi crossings is further verified by
fitting of the momentum distribution curve (MDC) at Fermi level acquired along the same direction as shown in Fig. 1
(c1). The two hole-like bands ($\alpha _{1}$ and $\alpha _{2}$) and two
electron-like bands ($\beta _{1}$ and $\beta _{2}$) can be also identified
from the peak positions of energy distribution curves (EDCs) as marked with color circles in Fig.1 (c3).
As shown in Fig. 2 (c1-d1), these bands hybridize leading to energy gaps 
(disconnected parts in the Fermi surface) and
reconstruction of the Fermi surface. 
In order to compare the band structure between the magnetic and non-magnetic
phase, we measured the ARPES data for CaFe$_2$As$_2$
deep below $T_N$ ($T=$12K) and slightly above $T_N$ ($T=$180K). 
The Fermi surface map near the zone center and the band dispersion map measured along a diagonal cut (blue arrow in Fig.1 (a)) are plotted in Fig. 2(e1) and Fig. 1(d3) for below $T_N$ and Fig. 2(e2) and Fig. 1(d4) for above $T_N$. The data were measured with 21.2 eV photons. 
The electron band ($\beta$) is observed only below $T_N$ (Fig. 1 (d3)), and it is absent above $T_N$ (Fig. 1 (d2)). The associated ``flower-like" shape of
Fermi surface changes to nearly circular as expected from band calculations in
non-magnetic phase.

\begin{figure}[tbp]
\includegraphics[width=3.35in]{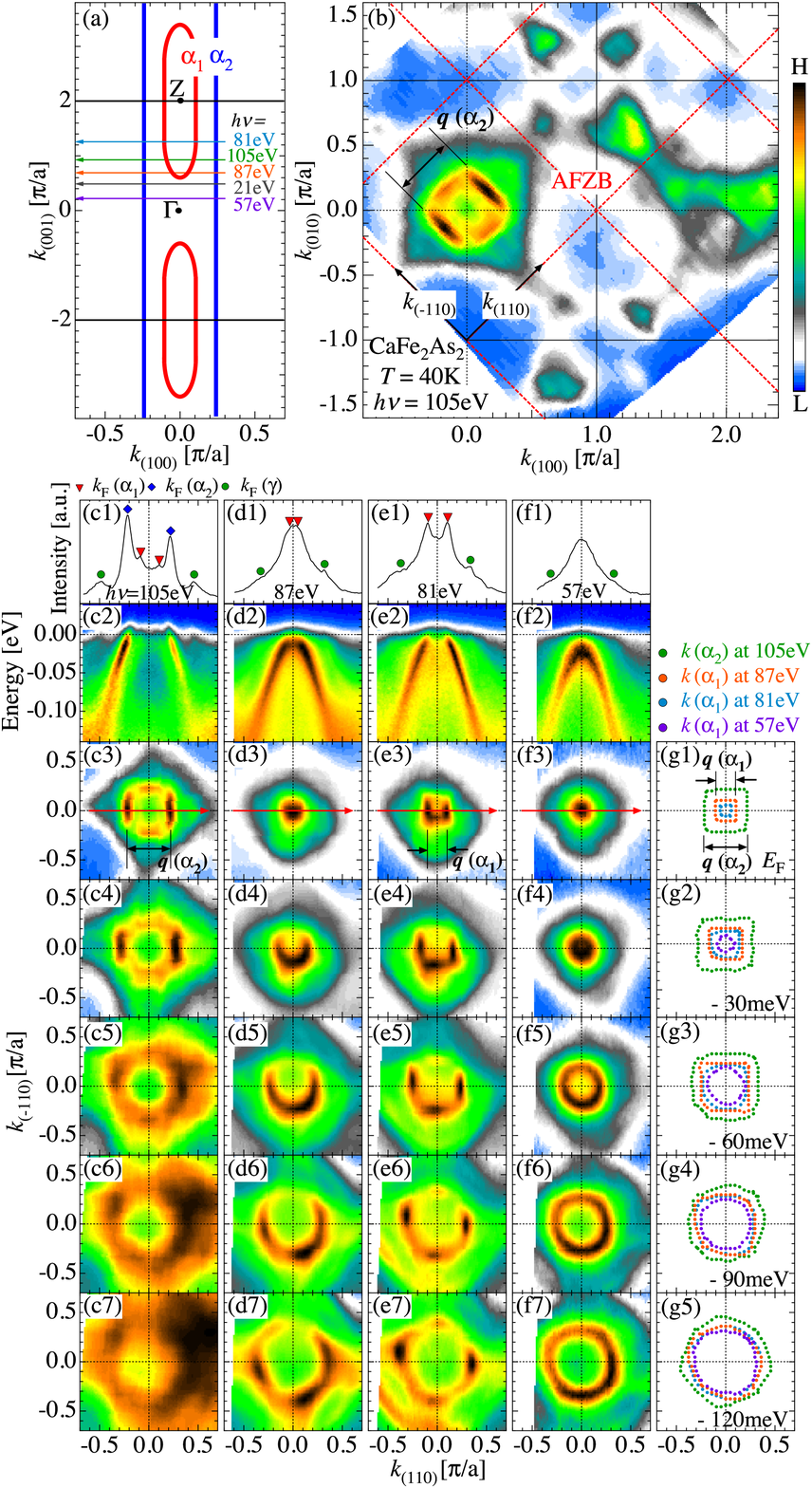}
\caption{(Color online) CaFe$_2$As$_2$ data at $T=40$K. (a) Schematic FSs along $k_{\mathrm{(001)}}$. (b) FS map along $a$-$b$ plane. (c1)-(f1) MDC of (c2)-(f2) at the Fermi level.
(c2)-(f2)  Energy dispersion map measured along a red arrow in (c3)-(f3). 
ARPES intensity about (c3)-(f3) Fermi level, (c4)-(f4) -30meV, (c5)-(f5)
-60meV, (c6)-(f6) -90meV, and (c7)-(f7) -120meV. (g1)-(g5) FS and energy contour plots extracted from (c-f). Only stronger peaks between $k_F(\alpha_1)$ and $k_F(\alpha_2)$ are plotted. 
Dimension arrows indicate a nesting vector $q(\alpha_1)$=$0.07(\pi/a, \pi/a)$ and $q(\alpha_2)$=$0.31(\pi/a, \pi/a)$  of hole-band $\alpha_1$ and $\alpha_2$, respectively.}
\label{fig3}
\end{figure}

We demonstrate that a simple model with the effect
of band back folding about the AFZB
reproduces many aspects of the data. Figure 2 (a1) shows a model of
two hole- and electron-pockets surrounding the zone center and the corner,
respectively, which are based on the band calculation for non-magnetic phase
\cite{Sknepnek09}. We assume that below $T_{N}$ the states around
$\Gamma$ (or Z) are coupled to the states around X by an interband staggered
mean field of magnitude $\Delta _{\mathrm{AF}}$
 via the mean field term in the Hamiltonian 
$
\sum\nolimits_{\mathbf{k},\sigma } {\sigma \Delta _{{\rm{AF}}} } \left( {d_{\mathbf{k},\sigma ,\rm{\Gamma (Z)}}^\dag  d_{\mathbf{k + Q},\sigma ,\rm{X}}^{}  + h.c.} \right)
$, where $\sigma$
refers to the spin and $Q$ is the magnetic ordering
vector. Upon increase of  $\Delta _{\mathrm{AF}}$ from 10meV
 to 50meV, the conduction bands are increasingly folded
back about the AFZB and the Fermi surfaces are reconstructed as shown in
Fig.2(a2-a4) including twinning effects\cite{Tanatar}. The degree of the reconstruction and hybridization gets stronger with increasing $\Delta _{\mathrm{AF}}$.
Eventually the FS shape becomes ``flower-like". We plot the
enlarged ARPES FS map of BaFe$_{2}$As$_{2}$ close to
$\Gamma$ and Z in Fig.2 (b1) and (c1), respectively, and the
extracted Fermi crossing points for the latter in Fig.2 (d1). The
experimentally observed ``flower-like" FS is well reproduced by
the simple model for $\Delta _{\mathrm{AF}}$ of 50meV (Fig. 2(a4)). The
reconstruction does not occur for $k_{z}$s where the
interband coupling via the momentum $Q$ is not efficient (Fig.
2(b1)). Figure 2 (b2) and (c2) shows the ARPES intensity close to $\Gamma $
and $Z$, respectively, slightly below the Fermi energy ($E=-20$meV). The
band positions for the latter are plotted in Fig. 2 (d2).
While the effect of band hybridization (``flower-like" shape) is clearly seen
even in this energy range around Z (Fig. 2(c2)), it is absent around $\Gamma$ (Fig. 2(b2))
with a round shape of the energy contour. The coupling between charge
carriers and magnetism is therefore strongest whenever efficient low energy
scattering  between the two sets of FS is kinematically possible, 
demonstrating the role of interband magnetic scattering in the iron pnictides. 
Previous ARPES studies \cite{Chang_Kz,Mannella} demonstrated strong dispersion along the $k_z$ direction.
This raised a question about the origin of magnetic order and pairing, since the presence of strong $k_z$ dispersion
would significantly limit in-plane $(\pi/a, \pi/a)$ nesting. 
By acquiring high quality data (Fig. 1(b)), we could see that, despite strong 3D-like dispersion, the Fermi
surface has a long ($\sim 1.7\pi /c$) parallel segments along the $k_z$.  The strong FS reconstruction discussed above is most prominent in this $k_z$ range. It might be essential for nesting conditions leading to the emergence of the $(\pi/a, \pi/a)$ magnetic order and superconductivity. 

Perhaps the most interesting experimental observation in our data is an
incommensurate FS nesting (long parallel segments of FS in the $a-b$ plane seen in Fig. 2 (d1) for BaFe$_2$As$_2$ and (e1) for CaFe$_2$As$_2$), which doesn't appear in the result of the model calculation (Fig.2 (a4)). The nesting vector $q(\alpha_1)\simeq0.1(\pi/a, \pi/a)$ and $q(\alpha_2)\simeq0.3(\pi/a, \pi/a)$ for hole-band $\alpha_1$ and $\alpha_2$, respectively, does not correspond to any previously reported density wave states.  Therefore, we speculate that the gapless, nested Fermi surface may be a precursor of a failed density wave order such as the CDW predicted by renormalization group studies \cite{DHLEE,DHLEE2}. In order to investigate the $k_z$ variation of the FS nesting, we measured FS maps along $a-b$ for CaFe$_2$As$_2$ at various photon energies ($h\nu$s). 
Figure 3(b) shows a typical FS map measured over a wide Brillouin zone with 105eV photons. 
The reconstructed ``flower-like" FS and the significant nesting with $q(\alpha_2)\simeq0.3(\pi/a, \pi/a)$ are clearly observed. In Fig. 3 (c3)-(f3), we plot FS maps focused on the zone center ($\Gamma$(Z)) measured at various $h\nu$s. The $k_z$s corresponding to these photon energies are indicated in Fig. 3 (a) with the 1st Brillouin zone and schematic FSs of hole-bands ($\alpha_1$ and $\alpha_2$). 
We note that the intensity of several bands ($\alpha_1$, $\alpha_2$, and $\gamma$ \cite{Chang_Kz}) dramatically changes with photon energy, and the electron bands ($\beta_1$ and $\beta_2$) are absent due to 
 matrix element effect. These are confirmed in the MDCs at the Fermi level shown in Fig. 3 (c1-f1) - the $\alpha_2$ is most intense at 105 eV, while the $\alpha_1$ is for 87eV, 81eV and 57eV photons. 
We also plot energy contour intensities in (c4)-(f4), (c5)-(f5), and (c6)-(f6) ($E=$-30meV, -60meV, -90meV and -120meV, respectively), which help us to understand how these bands behave with the binding energy. 
We found that the signature of band reconstruction is absent at specific $k_z$ where the band doesn't cross $E_F$ (Fig.3 (f1)-(f7)), and quickly evolved when the Fermi pockets starts appearing.
This indicates that the band hybridization in the magnetic phase occurs only when the Fermi pockets overlap with folding-back bands and the gap opens with the consequent energy gain at the overlapped segments (seen as disconnected parts in Fig.1(c1) and (d1)). 
The parallel segments in the energy contours are significant up to $E\simeq$-100meV. This binding energy seems to agree with the energy bottom of electron-band back folded from the zone corner (see Fig. 1 (d3)). 
This indicates that  the failed density wave order is tied to the existence of magnetic long range order, and suggest a
strong coupling between the staggered magnetization and other density wave order parameters in the iron pnictides.

In conclusion, we find strong band back folding and hybridization effects
that lead to a reconstruction of the Fermi surface (FS) below the magnetic
ordering temperature ($T_{N}$). The round hole-pockets observed above $T_{N}$ evolve
into a square shape surrounded by ``flower petal" electron pockets below $
T_{N}$. The FS data can be reasonably understood within a simple model of
magnetic ordering. More interestingly, we find long parallel segments of
the Fermi surface in the $a-b$ plane, which is not expected by the simple model.
We speculate that these nested parts of the Fermi surface may be a precursor of
a failed density wave order such as the CDW predicted by renormalization
group studies \cite{DHLEE,DHLEE2}. These findings demonstrate
the complexity of these newly discovered materials and form the foundation of
understanding the electronic properties that lead to high temperature
superconductivity in the carrier-doped phase.

We thank A. Kreyssig, A. I. Goldman and B. N. Harmon for insightful
discussions. ALS is operated by the US DOE under Contract No.
DE-AC03-76SF00098. Ames Laboratory was supported by the US DOE
- Basic Energy Sciences under Contract No. DE-AC02-07CH11358.


\end{document}